\documentclass[12pt]{iopart}
\usepackage[utf8]{inputenc}
\usepackage[T1]{fontenc}
\usepackage{ae,aecompl} 
\usepackage{graphicx}
\usepackage{color}
\usepackage{amssymb}
\usepackage{latexsym}
\usepackage{wasysym}
\usepackage{psfrag}
\usepackage{ifthen}
\usepackage[citecolor=blue,colorlinks=true]{hyperref}
\usepackage{longtable}
\usepackage{float}
\usepackage{lineno}
\usepackage{units}
\usepackage{multirow}
\usepackage{ulem}

\newcommand{\msun}{\rm {M\textsubscript{\(\odot\)}}}

\newcommand{\dlumi}{D_{\rm{L}}}

\begin{document}

%\begin{center}
% Version 2.1
%\end{center}

% ======================
% TITLE AND ABSTRACT
% ======================
\title[Searches for Compact Binary Coalescence Events using CNNs in LIGO/Virgo]{Searches for Compact Binary Coalescence Events Using Neural Networks in LIGO/Virgo Third Observation Period}

\author{A.~Men\'endez-V\'azquez$^1$, M.~Andrés-Carcasona$^1$, M.~Mart\'inez$^{1,2}$, Ll.~M.~Mir$^{1,*}$}

\address{$^1$Institut de F\'isica d’Altes Energies (IFAE), The Barcelona Institute of Science and
Technology, Campus UAB, E-08193 (Barcelona), Spain}
\address{$^2$Instituci\'o Catalana de Recerca i Estudis Avançats (ICREA), E-08010 (Barcelona), Spain}

\address{$*$ Corresponding author: mir\@ifae.es}

\date{\today}

% ===================
% ABSTRACT
% ===================
\begin{abstract}

We present the results on the search for the coalescence of compact binary mergers
using convolutional neural networks  and 
the  LIGO/Virgo data for the O3 observation period. 
Two-dimensional images in time and frequency are used as input.  
The analysis is performed in three separate mass regions  covering the range for the masses in the  binary system from 0.2~$\msun$ to 100~$\msun$, excluding very asymmetric mass configurations. We explore neural networks trained with input information from pairs of interferometers  or all three 
interferometers together,  concluding that the use of the maximum information available  
leads to an improved performance. A scan over the O3 data set, using the 
convolutional neural networks, is performed with different false rate thresholds for claiming detection of at most one event per year or at most one event per week. The latter would correspond to a loose online selection still leading to affordable false alarm rates. The efficiency of the neutral networks to detect the O3 catalog events is discussed. In 
the case of a false rate threshold of at most one event per week,  the scan leads to the detection of about $50\%$ of the O3 catalog events. Once the
search is limited to the catalog events within the mass range used for neural networks training, the detection efficiency increases up to 70$\%$. 
Further improvement in search efficiency, 
using the same type of algorithms, 
will require the implementation of new criteria to suppress the remaining major background sources.

\end{abstract}

\maketitle

% ===================
%  INTRODUCTION
% ===================
\section{Introduction}
\label{sec:intro}
The use of artificial intelligence tools to search for gravitational wave (GW) event candidates in the LIGO-Virgo data remains a very active field of research. In the case of GW events from compact binary coalescence (CBC) of black holes (BH) and/or neutron stars (NS),  this is mostly motivated by the fact that the traditional approach,  based on the extraction of the GW signal out of a much larger noise in the data using matched-filtering techniques and huge banks of GW waveform templates, is very demanding in terms of computing resources. In particular, the presence of a distinct chirp-like shape in the CBC events, 
when represented in spectrograms showing the signal in  frequency-time domain, 
makes the use of a convolutional neural network (CNN)  a valid alternative suitable for GW detection~\cite{PhysRevLett.120.141103,george2017deep,Gebhard_2019,George_2018,Menendez-Vazquez:2020khz,PhysRevD.107.082003}. 
In addition, 
the use of CNNs has been explored in the past to distinguish between families of glitches~\cite{Razzano_2018,Biswas_2013,Cavaglia_2019}
and to determine the physical parameters of GW events~\cite{10.1093/mnras/stad3448}.

In this paper, we follow closely the analysis procedure  in~\cite{Menendez-Vazquez:2020khz,PhysRevD.107.082003} to search for CBC events in the data from the third LIGO-Virgo-KAGRA observation run (O3). 
Compared to Ref.~\cite{Menendez-Vazquez:2020khz}, which analysed the O2 data, a number of improvements are implemented.  New signal regions in the masses of the binary system, covering the whole mass range between 0.2~$\msun$  and 100~$\msun$, are introduced, and a new approach in determining the CNN working point for signal discrimination, based on the computed false alarm rate (FAR) for the selected candidates, is employed.

% ===================
% DATA ANALYSIS
% ===================
\section{Data preparation}
\label{sc:dp}

The study uses the O3 data~\cite{KAGRA:2023pio} from LIGO-Livingston (L1), LIGO-Hanford (H1) and Virgo (V1) interferometers with 4096 Hz sampling rate,
obtained from the Gravitational Wave Open Science Center~\cite{gwosc}.
The data cover from 1 April 2019 1500 UTC to 27 March 2020 1700 UTC, divided in two periods, denoted as O3a and O3b, separated by a commissioning period of one month in October 2019. 
After applying data quality requirements dealing with the understanding of the interferometer stationary noise budget 
as well as the identification and suppression of glitches and spectral noise contributions~\cite{Davis_2021, Acernese_2023}, 
the sample includes a total of 155 days of combined H1-L1-V1 data.
A fraction of these data is used to construct background and background plus injected signal images for the purpose of training the CNNs. An additional veto of times containing O3 GW events, 
as indicated in the GWTC-3 catalog~\cite{2021PhRvX..11b1053A, PhysRevD.109.022001, PhysRevX.13.041039}, 
is included to avoid contamination of the background sample. The signals are generated using the PyCBC package~\cite{Usman2016TheCoalescence,nitz2017detecting,nitz2020gwastro} and the waveform approximant \texttt{IMRPhenomPv2}~\cite{PhenomPv3}. Signals are combined with the background data from the different interferometers after taking into account the proper relative orientations, times of arrival and antenna factors. 

\begin{figure}[htb]
\begin{center}
\includegraphics[width=0.495\textwidth, height=6 cm]{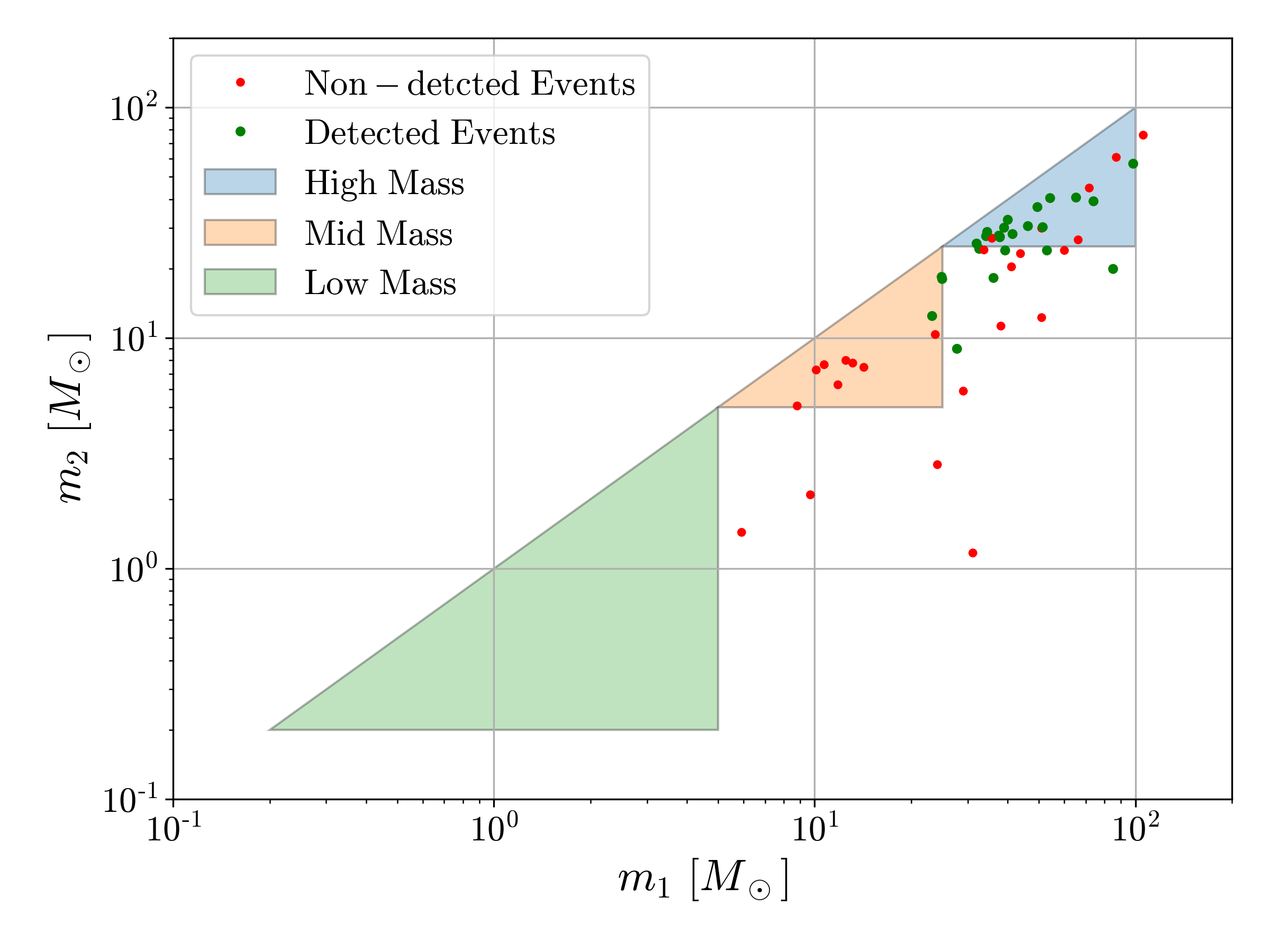}
\end{center}
\caption{\small
Definition of the search regions in the $m_1 - m_2$ mass plane. 
The dots indicate the masses of the binary systems corresponding to the GWTC-3 
catalog~\cite{2021PhRvX..11b1053A, PhysRevD.109.022001, PhysRevX.13.041039},
in green those selected by the CNNs in this work and in red those which are not.
}
\label{fig:regions}
\end{figure}

As presented in Figure~\ref{fig:regions}, 
the mass range of the binary system is separated in three regions with increasing masses of the binary components 
$m_1$ and $m_2$, with $m_1 > m_2$.  
As in Ref.~\cite{Menendez-Vazquez:2020khz},
the low-mass region covers the mass range between 0.2 - 5.0 $\msun$ and
a corresponding luminosity distance, $\dlumi$, is limited to 100~Mpc; 
and the high-mass region includes the mass range between 25 and 100~$\msun$ with $\dlumi$ between 100~Mpc to 1400~Mpc. 
This is now complemented with an intermediate-mass region covering the mass range between 5 and 25~$\msun$ and $\dlumi$ in the range 1 to 1000~Mpc. 
The different limitations in $\dlumi$ take into account the O3 observed $\dlumi$ distribution in the data and the expected sensitivity at very large distances.  
In the CNN training process, the configurations with $m_2/m_1 < 0.05$ are excluded. 
The latter corresponds to very asymmetric mass configurations for which a dedicated CNN search~\cite{PhysRevD.107.082003} has been performed separately.  
Figure~\ref{fig:regions} also collects the masses corresponding to the GWTC-3 catalog events 
and indicates whether the events are finally detected by this work, as discussed below.   
The generated event parameters are uniformly distributed,  
as indicated in Table~\ref{tab:params}. 

\begin{table}[htb]
\begin{center}
\begin{tabular}{c|ccc}
\hline
\multirow{2}{5em}{Parameters} & \multicolumn{3}{c}{Mass range}  \\  \cline{2-4}
                              & Low & Intermediate & High  \\ 
\hline
$m_1$, $m_2$ {[}\msun{]}    & {[}0.2, 5{]}             & {[}5, 25{]}               & {[}25, 100{]} \\
$\dlumi$ {[}Mpc{]}          & {[}1, 100{]}             & {[}1, 1000{]}             & {[}100, 1400{]} \\
$\theta_{JN}$ {[}rad{]}     & {[}0, $\frac{\pi}{2}${]} & {[}0, $\frac{\pi}{2}${]}  & {[}0, $\frac{\pi}{2}${]} \\
$\psi$ {[}rad{]}            & {[}0, $\pi${]}           & {[}0, $\pi${]}            & {[}0, $\pi${]} \\
$\alpha$ {[}rad{]}          & {[}0, 2$\pi${]}          & {[}0, 2$\pi${]}           & {[}0, 2$\pi${]} \\
$\delta$ {[}rad{]}          & {[}0, $\pi${]}           & {[}0, $\pi${]}            & {[}0, $\pi${]} \\ 
$\mathcal{M}_c$ {[}\msun{]} & {[}0.17, 4.35{]}         & {[}4.35, 21.76{]}        & {[}21.76, 87.06{]} \\

\hline
\end{tabular}
\end{center}
\caption{\small
Signal generation parameters in each of the three signal regions. Here $m_1$ and $m_2$ are the black hole masses in the binary system, computed in the detector frame, with $m_1 > m_2$; $D_L$ is the luminosity distance;  
$\theta_{JN}$ denotes the inclination of the orbit with respect to the line of sight; $\psi$  is the polarization of the GW;  and $\alpha$ and $\delta$ are the right ascension and declination, respectively.
Finally, $\mathcal{M}_c$ shows the corresponding chirp mass range, 
defined in section~\ref{sec:injection}, 
for each signal region.
}
\label{tab:params}
\end{table}

In order to control the duration of the signals, a low frequency threshold of 80~Hz is applied to the signals in the low- and intermediate-mass regions. For the high-mass region, this threshold is reduced to 25~Hz.  The signals duration is limited to 
five seconds counting backwards from the merger time to remove low frequency components that might confuse the neural network.
The generated signals are randomly placed within five second windows of data from each interferometer before being processed. Background and background plus signal segments are whitened following the same prescription as in Ref.~\cite{LIGOScientific:2019hgc}. The whitened segments are used to produce spectrograms using $Q$-transforms~\cite{QTransf} with 400 bins in time and 100 bins in frequency. Finally, the images are processed such that their content has zero average and variance equal to one, following the same prescription as in Ref.~\cite{Menendez-Vazquez:2020khz}.

% ===================
% NN definition and training
% ===================
\section{Neural network definition and training}
\label{sc:nn}
\begin{figure}[H]
\begin{center}
\includegraphics[width=0.495\textwidth, height=6 cm]{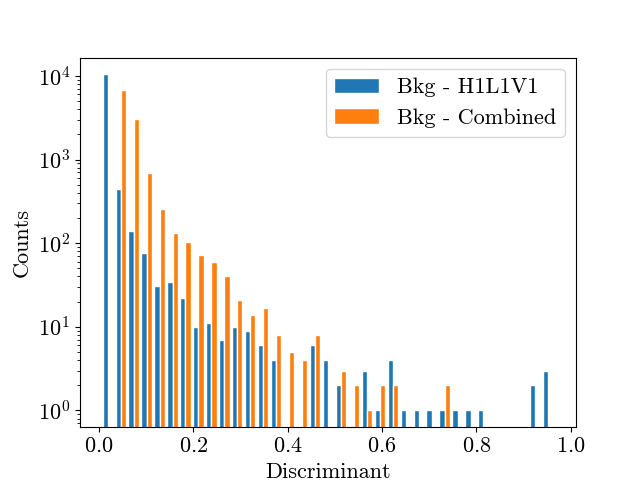}
\includegraphics[width=0.495\textwidth, height=6 cm]{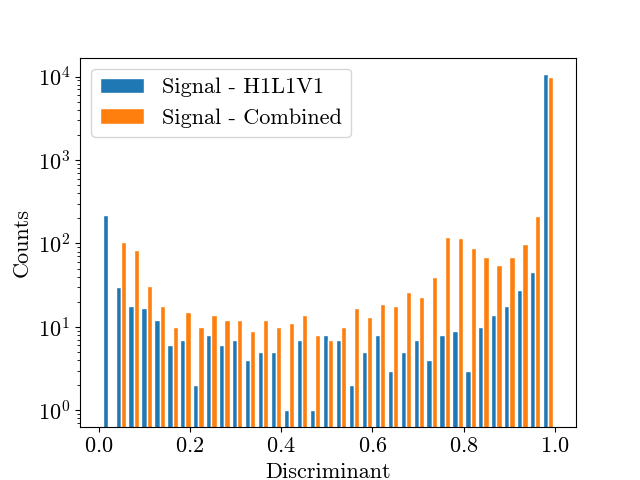}
\end{center}
\caption{\small
Comparison of the neural network discriminants for (left) background data and (right) simulated signals using the arithmetic mean of all the CNN outputs and only the H1-L1-V1 combination.
}
\label{fig:Bkg_comp}
\end{figure}

As in previous studies, we adopted a deep CNN {\it{ResNet-50}} with a 50-layer architecture, as described in~\cite{DBLP:journals/corr/HeZRS15}. 
Binary cross-entropy was used as a loss function along with Adam as the optimizer~\cite{Kingma2014Adam:Optimization}. 
The CNN  training was based on the first O3a data period. 
No attempt  was  made to carry out a new training process for the O3b data since the sensitivity of the instruments remained stable during the whole O3 observation period. 
The dataset consists of 250,000 images, 
of which 115,000 were used for training, 
120,000 for testing and 15,000 for validation, 
evenly distributed into background-only and background with a signal injected.
A total of four CNNs per mass range were trained covering all the combinations of interferometer inputs: H1-L1, H1-V1, L1-V1 and the triple combination, H1-L1-V1. The use of a single interferometer as input to the CNNs was discarded due to the lower performance already observed in the past compared to the use of multiple inputs. Each neural network is trained for up to 12 epochs and the one with the lowest error over the validation set was selected. 
In all cases the training showed a stable behaviour. 
Keras was used to build and train the CNN with the TensorFlow backend 
and its implementation on GPUs~\cite{abadi2016tensorflow}.
Training typically takes two hours per CNN, 
and overfitting is avoided by checking the CNN performance on the validation set after each epoch.
A further improvement of the global sensitivity is achieved by combining the outputs of the separate CNNs into a global discriminant. Such combination provides an additional tool for suppressing glitches in the data affecting independently the interferometers and in different time stamps. As in the case of Ref.~\cite{PhysRevD.107.082003}, a simple average of the H1-L1-V1, H1-L1, L1- V1, and H1-V1 CNN outputs has been considered. As shown in Figure~\ref{fig:Bkg_comp}, this translates into a better separation of the background and signal and a decrease in the amount of false positives detected by the neural networks. 

\begin{figure}[H]
\begin{center}
\includegraphics[width=0.495\textwidth,height=6 cm]{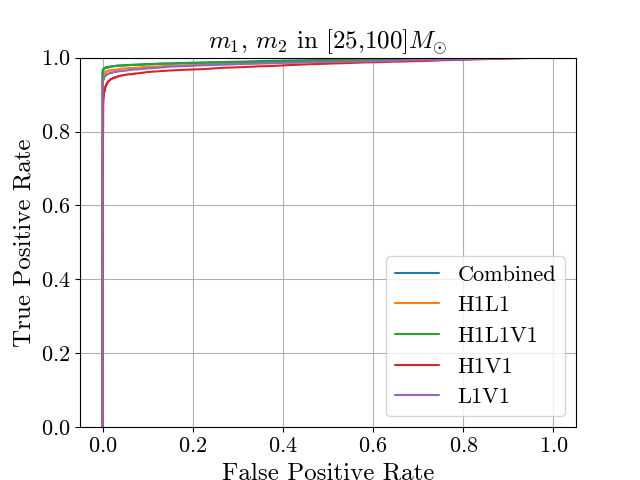}
\includegraphics[width=0.495\textwidth,height=6 cm]{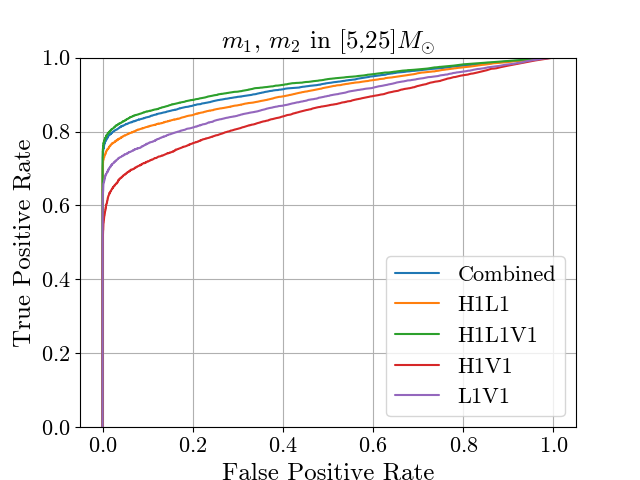}
\includegraphics[width=0.495\textwidth,height=6 cm]{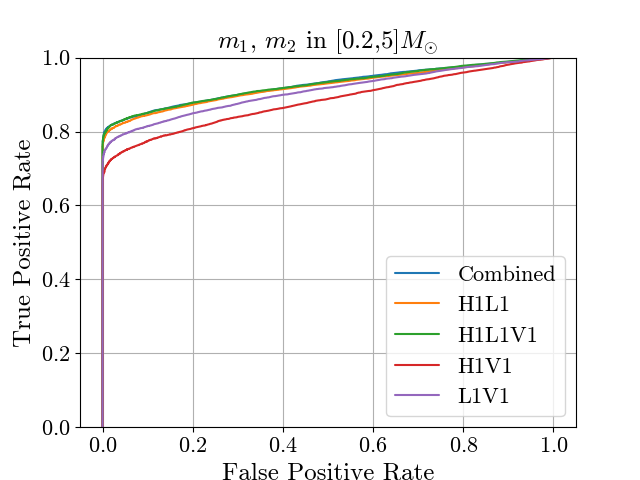}
\end{center}
\caption{\small
The ROC curves  for the different CNNs including different inputs and for (top-left) high-mass; (top-right) intermediate-mass; and (bottom) low-mass regions.
}
\label{fig:roc}
\end{figure}

The performance of the neural networks is presented in Figure~\ref{fig:roc} in terms of the receiver operating characteristic (ROC)  curves for the separate CNNs and their combination, representing the true positive (TP) versus the false positive (FP) rates. The best performance is achieved by the high mass neural network, reaching high TP values at low FP. This is to a large extent expected since the high-mass signals are generally louder and more visible in the instruments.  

As already mentioned, for each image the CNN discriminant is associated to a FAR value,  and the claim for observing a CBC candidate is determined by a predefined FAR threshold.
The FAR is estimated as a function of the discriminant of the CNN and is defined as ${\rm{FAR (\eta)}} = N(\eta)/T$, where $\eta\in[0,1]$ is the CNN discriminant, $N(\eta)$ is the number of events with a CNN discriminant above or equal to $\eta$ and $T$ the period of time analysed. In order to effectively increase the time considered in the FAR calculation, thus reaching very low FAR values, the time slide technique \cite{Abbott2005SearchStars} is used. 
This allows accumulating $O(10^9)$ images (of 5~s of duration each) and accessing FAR values down to $1/152.6$~years${}^{-1}$.  
Using this method,  we will consider a possible CBC detection when the combined CNN discrimination has an associated FAR value lower than either one event per year 
or  one event per week, where the latter is used to explore a looser selection  more adequate for an online implementation of the algorithm, while maintaining the false rate at a tenable level.

\section{Injection tests}
\label{sec:injection}

Signals with known parameters and signal-to-noise ratios ($\rho$) are injected in real data to understand the performance of the neural networks. The injected signals follow the same distribution as the ones used during the training but sampled uniformly in comoving volume, in agreement with the observed distribution of galaxies in the universe, and the masses are converted into the source frame assuming Planck15~\cite{Ade:2015xua} cosmology from the Astropy~\cite{astropy:2018} Python package.

\begin{figure}[H]
\begin{center}
\includegraphics[width=0.495\textwidth]{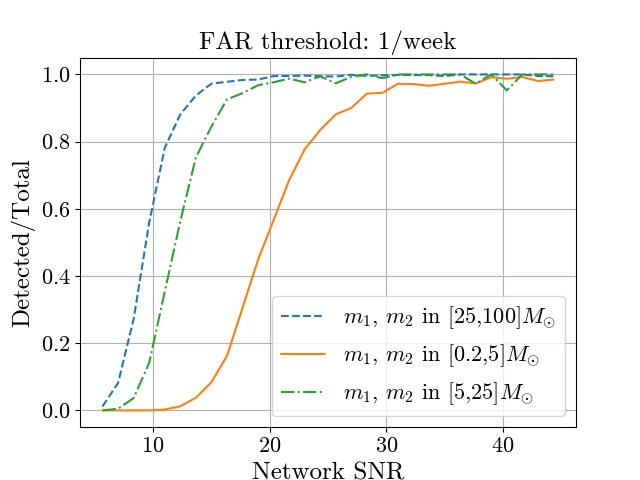}
\includegraphics[width=0.495\textwidth]{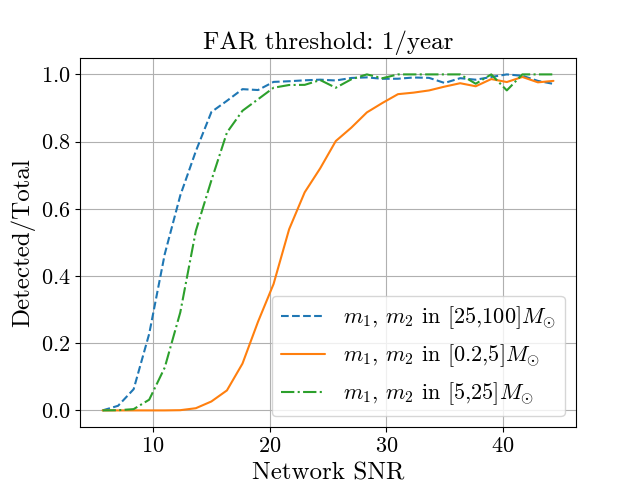}
\end{center}
\caption{\small
Efficiency as a function of the network SNR corresponding to the combination of all the NNs in each of the mass ranges and for a FAR threshold of (left) one event per week  and (right) one event per year.
}
\label{fig:lowinject1}
\end{figure}

For each GW signal, the  value for $\rho$ is computed following the 
prescription in Ref.~\cite{PhysRevLett.120.141103} solving the integral 

\begin{equation}
    \rho^2 = \int_{f_{\rm{min}}}^{f_{\rm{max}}}  df \ |h(f)^2|/S_n(f), 
\end{equation}
\noindent

\noindent in the frequency domain $(f)$, where $|h(f)^2|$ denotes the signal and $S_n(f)$ 
is the power spectral density of the background. 
We define the network signal-to-noise ratio (SNR), $\mathrm{SNR_{net}}$, as

\begin{equation}
    \mathrm{SNR_{net}} = \sum_{i}{\rho_i^2},
\end{equation}

\noindent where $i$ is an index that runs over the different interferometers. 
A total of 32,000 injections per NN were performed. 
Figure~\ref{fig:lowinject1} shows the fraction of GW signals identified by the CNNs as a function of $\mathrm{SNR_{net}}$ for the combination of all the neural networks  in the different mass regions and for the two FAR thresholds considered.  
As expected, a looser FAR requirement translates into an improved detection efficiency at a given SNR.  

\noindent
In the case of a FAR threshold of one event per week, both the high-mass and intermediate-mass neural networks show good efficiency and are almost fully efficient at network SNRs above 20. The low-mass range network instead only becomes fully efficient at network SNRs above 39. This is attributed to the difficulty in detecting low-mass signals not always visible in the images.  
The same tendency is observed for a FAR threshold of one event per year although the efficiency curve shows a more moderate increase with increasing SNR.
Table~\ref{tab:rho} collects the values of $\mathrm{SNR_{net}}$ for different efficiency values. 

\begin{table}[H]
\begin{center}
\begin{tabular}{l c c c}\hline
\multicolumn{4}{c}{FAR threshold of 1 event per week}\\ \hline
 Mass Range& $\mathrm{SNR_{net}} (50\%)$ & $\mathrm{SNR_{net}} (80\%)$ & $\mathrm{SNR_{net}} (99\%)$\\ 
 \hline
High& 10 & 12 &  20\\
Intermediate & 12 & 15 &  24\\
Low  & 20 & 24 &  39 \\ \hline
\multicolumn{4}{c}{FAR threshold of 1 event per year}\\ \hline
 Mass Range& $\mathrm{SNR_{net}} (50\%)$ & $\mathrm{SNR_{net}} (80\%)$ & $\mathrm{SNR_{net}} (99\%)$\\ 
  \hline
 High& 12 & 15 & 28 \\
Intermediate & 14 & 16 & 28 \\
Low  & 21 & 25 & 41  \\ \hline
\end{tabular}
\caption{\small 
Values of network SNR at given detection efficiencies for the different 
CNNs and a FAR threshold of 1 event per week and 1 event per year (see body of the text).
}
\label{tab:rho}
\end{center}
\end{table}

\begin{figure}[H]
\begin{center}
\includegraphics[width=0.495\textwidth,height=6cm]{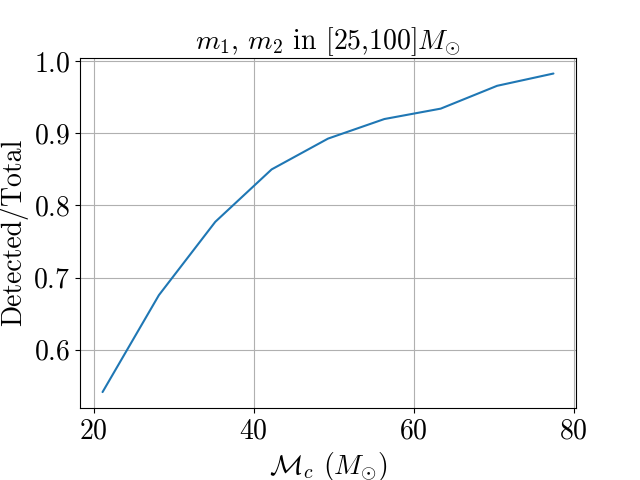}
\includegraphics[width=0.495\textwidth,height=6cm]{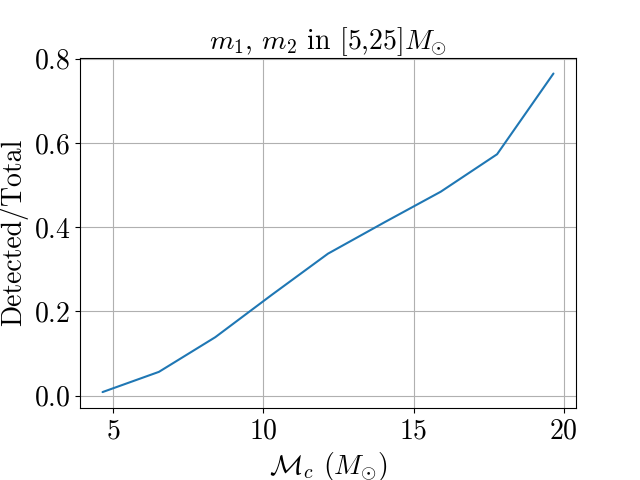}
\includegraphics[width=0.495\textwidth,height=6cm]{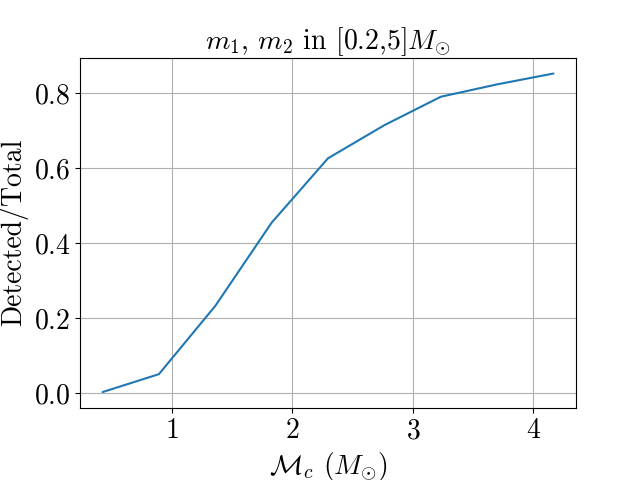}
\end{center}
\caption{\small
Detection efficiency of the CNNs as a function of the chirp mass corresponding to a FAR threshold of one event per week in (top-left) the high-mass (top-right) the intermediate-mass and (bottom) the low-mass regions. 
}
\label{fig:eff_chirpM}
\end{figure}

Figure~\ref{fig:eff_chirpM} presents the efficiency for event detection as a function of the chirp mass, 
defined as 
$\mathcal{M}_c \equiv \frac{(m_1 m_2)^{3/5}}{(m_1 + m_2)^{1/5}}$, 
separately for each mass range. 
A FAR threshold of one event per week is used in this case.  In general, the detection efficiency increases with increasing the chirp mass. In the case of the high-mass range, the CNN efficiency increases from  50$\%$ at 20~$\msun$ and $90\%$ at 50~$\msun$ to 98$\%$ at 75~$\msun$.  In the intermediate-mass range, the CNN shows a marginal efficiency at 5~$\msun$ which increases almost linearly reaching a value of 70$\%$ at 19~$\msun$. Finally, in the low-mass range, the CNN efficiency increases from  10$\%$ at 1~$\msun$ and 50$\%$ at 2~$\msun$ to 85$\%$ at 4~$\msun$. As expected, the CNN performance is limited at low chirp mass. This indicates that the NNs easily recognize sharp features in higher mass signals and somehow fail to detect low tails in the spectrograms associated with low-mass GW signals.  

% ======================
% RESULTS
% ======================
\section{Results}
\label{sec:results}
\begin{figure}[H]
\begin{center}
\includegraphics[width=0.76\textwidth]{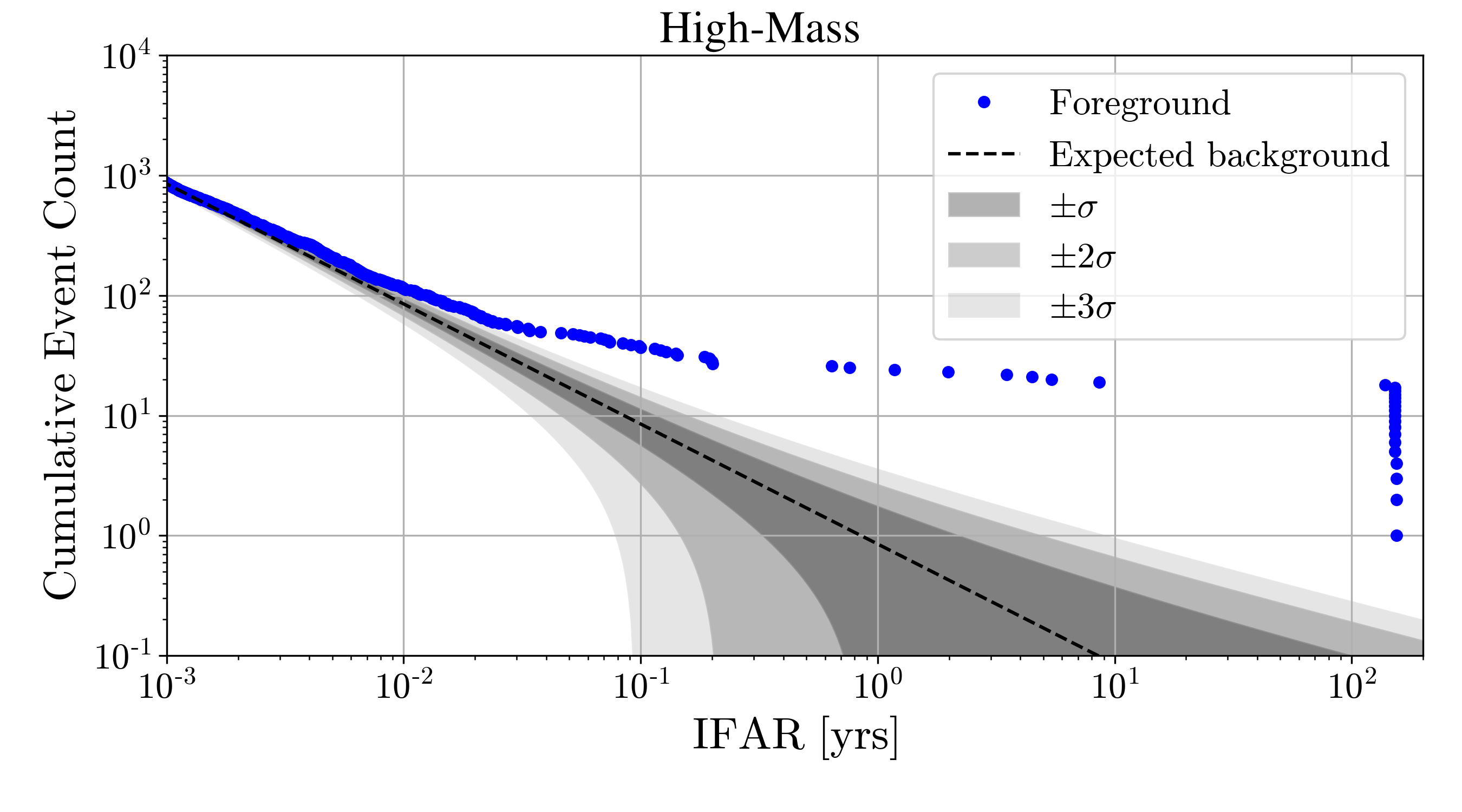} \\
\includegraphics[width=0.76\textwidth]{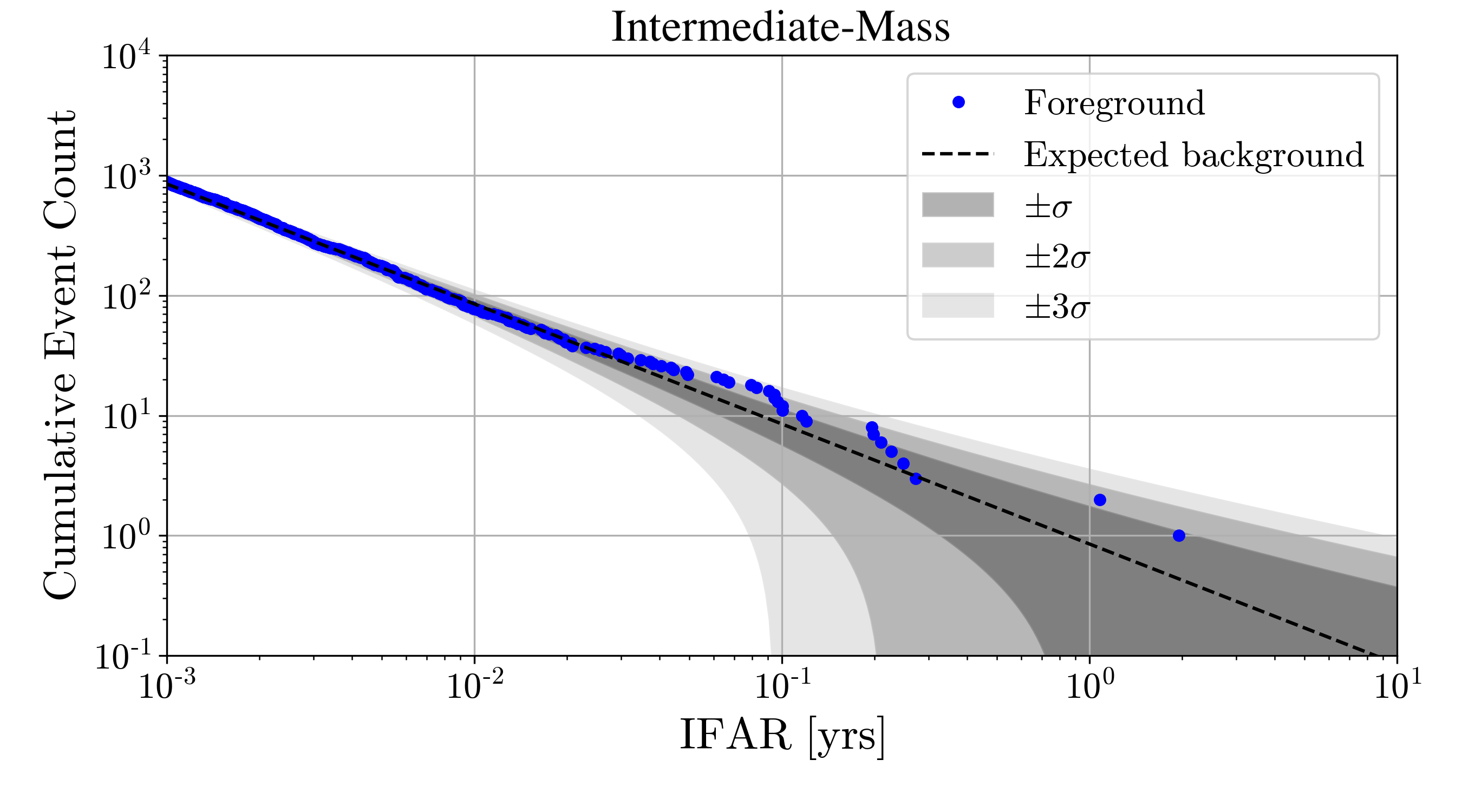} \\
\includegraphics[width=0.76\textwidth]{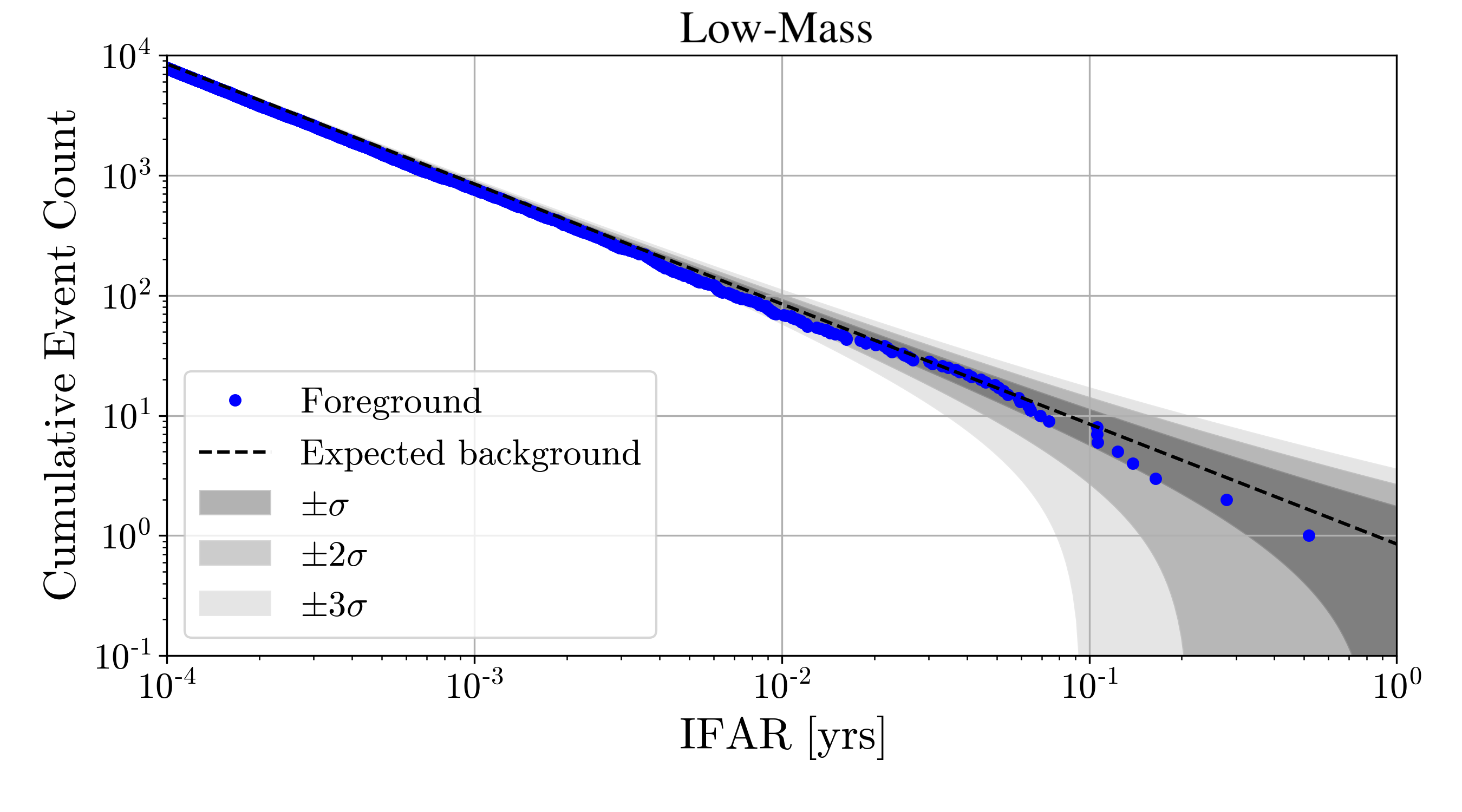}
\end{center}
\caption{\small
The measured IFAR distributions (dots) for O3 data for the different CNNs in the (top) high-mass, 
(middle) intermediate-mass and (bottom) low-mass regions.  
The results are compared to the background only expectations (dashed lines) and the 
corresponding $\pm$1$\sigma$, $\pm$2$\sigma$ and $\pm$3$\sigma$ uncertainty bands.
}
\label{fig:O3_scan}
\end{figure}

The CNN global discriminating outputs in the different mass ranges, 
defined as the average of the corresponding H1-L1-V1,  H1-L1, L1-V1, and H1-V1 CNN outputs, 
are used to search for CBC signals. 
Here we limit ourselves to the analysis of the data for which all the three interferometers were declared in science mode. 
A slicing window of five seconds duration was used in steps of 2.5 seconds 
(leading to a 50$\%$ overlap between consecutive images) for each of the interferometers.  
A scan over the data using different global discriminating values in the range between 0 and 1 is performed.  
In each case, the corresponding FAR is computed. 
The computation time for the entire O3 analysis has been of the order of 2000 CPU hours per search 
(on an Intel Xenon CPU E5-2680 v4 \@ 2.4 GHz),
which is a significant improvement over the CPU time typically required for known matched-filtering pipelines.
Figure~\ref{fig:O3_scan} shows the resulting  inverse FAR (IFAR) distributions in the separate mass ranges, in units of years, compared to the expected yields of noise events following Poisson probability distributions. 

In the high-mass range an excess of detections above the only-noise prediction is observed.  
This excess is attributed to the detection of CBC events present in the GWTC-3 catalog. 
It vanishes once the CBC events are excluded from the data. 
No excess of events above the only-noise prediction is observed in the other mass regions.

According to the GWTC-3 catalog, 
the data analysed in this paper include 50 O3 events in which all three interferometers are in science mode.
Using a FAR threshold of one event per week the HM CNN detects 26 of those events, 
corresponding to an efficiency of~$\approx 50\%$.  
However, 
only 31 of the 50 events have chirp masses within the CNN training range,  
and of those 22 are detected, 
corresponding to a detection efficiency of~$\approx 70\%$, 
close to the value anticipated by the signal injection studies.  

\begin{figure}[H]
\begin{center}
\includegraphics[width=0.495\textwidth]{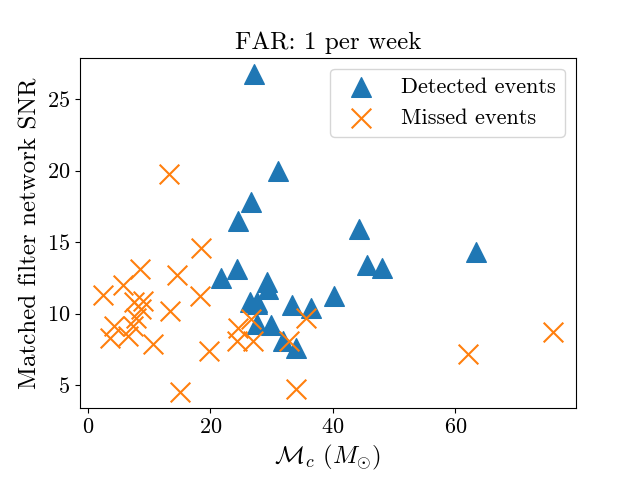}
\includegraphics[width=0.495\textwidth]{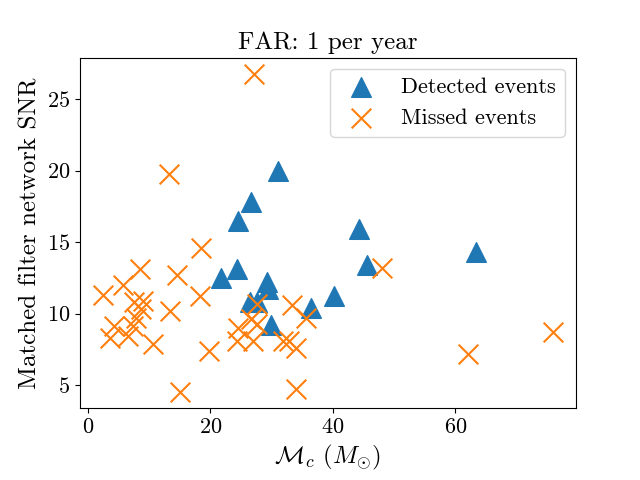}\\
\end{center}
\caption{\small 
The network SNR  
versus the chirp mass for the O3 GWTC-3 catalog events~\cite{2021PhRvX..11b1053A, PhysRevD.109.022001, PhysRevX.13.041039} 
with all the three detectors were in science mode. 
The results are shown for FAR values of one event per week (left) and one event per year (right).
The crosses indicate events missed by the CNN and the triangles denote CNN detected events.
}
\label{fig:2d1y}
\end{figure}

Figure~\ref{fig:2d1y} presents the correlation between the network SNR and the chirp mass.  
As expected,
given the low detection efficiency of the CNNs for low SNR events observed in Figure~\ref{fig:lowinject1},
detections are clustered at large SNR and large masses.
With the current model and for large values of the discriminant, 
small discriminant variations translate into large variations in FAR,
and discriminant values around 0.95 are necessary to reduce the FAR to one event per year.
Consequently, several events with values very close to the threshold are not detected.
Finally, \ref{sec:appendix} collects the numerical results for the subset of O3 CBC events for which all the three detectors were in science mode. 
Improving the detection efficiency for a given FAR rate beyond the observed results, 
using CNNs and two-dimensional images, 
would require a full characterisation of the major remaining noise sources and the implementation 
of new criteria for their suppression before the CNN algorithms are applied,
which is beyond the scope of this study.  

% ======================
% SUMMARY
% ======================
\section{Summary}
\label{sec:sum}
We present an update on the search for the coalescence of compact binary mergers
using LIGO/Virgo data and convolutional neural networks based on two-dimensional images in time and frequency as input. 
The analysis is performed in three separate mass regions covering the mass range 0.2~$\msun$ to 100~$\msun$. 
A scan over the O3 LIGO/Virgo data set, 
using a false rate threshold for claiming detection of at most one event per week,  
leads to the detection of about $50\%$ of the O3 catalog events. 
Once the search is restricted to the catalog events within the mass range used for neural networks training,  
the detection efficiency increases up to about 70$\%$. 
The search is limited by a large rate of false positives originating from background. 
Further improvement in search efficiency, 
using the same type of algorithms, 
will require the implementation of new criteria to suppress the remaining major background sources.

% ======================
% Acknowledgments
% ======================
\section*{Acknowledgements}

This research has made use of data or software obtained from the Gravitational Wave Open Science Center (gwosc.org), 
a service of the LIGO Scientific Collaboration, the Virgo Collaboration, and KAGRA. 
This material is based upon work supported by NSF's LIGO Laboratory which is a major facility fully funded by the National Science Foundation, as well as the Science and Technology Facilities Council (STFC) of the United Kingdom, the Max-Planck-Society (MPS), and the State of Niedersachsen/Germany for support of the construction of Advanced LIGO and construction and operation of the GEO600 detector. 
Additional support for Advanced LIGO was provided by the Australian Research Council. 
Virgo is funded, through the European Gravitational Observatory (EGO), by the French Centre National de Recherche Scientifique (CNRS), the Italian Istituto Nazionale di Fisica Nucleare (INFN) and the Dutch Nikhef, with contributions by institutions from Belgium, Germany, Greece, Hungary, Ireland, Japan, Monaco, Poland, Portugal, Spain. 
KAGRA is supported by Ministry of Education, Culture, Sports, Science and Technology (MEXT), Japan Society for the Promotion of Science (JSPS) in Japan; National Research Foundation (NRF) and Ministry of Science and ICT (MSIT) in Korea; Academia Sinica (AS) and National Science and Technology Council (NSTC) in Taiwan.
The authors thankfully acknowledge the computational resources provided by PIC and MinoTauro, 
and the technical support provided by Barcelona Supercomputing Center (RES-FI-2021-3-0020). 
This paper has been given LIGO DCC number LIGO-P2400028 and Virgo TDS number VIR-0065A-24.
This work is partially supported by the Spanish MCIN/AEI/ 10.13039/501100011033 under the grants SEV-2016-0588, PGC2018-101858-B-I00, and PID2020-113701GB-I00 some of which include ERDF funds from the European Union. 
IFAE is partially funded by the CERCA program of the Generalitat de Catalunya. 
M.~A-C. is supported by the 2022 FI-00335 grant.
This work was carried out within the framework of the EU COST action No. CA17137.

% =====================
% BIBILIOGRAPHY
% =====================
\section*{References}
\bibliographystyle{unsrt}
\bibliography{bibfile}

\begin{thebibliography}{10}

\bibitem{PhysRevLett.120.141103}
H.~Gabbard, M.~Williams, F.~Hayes, and C.~Messenger.
\newblock Matching matched filtering with deep networks for gravitational-wave
  astronomy.
\newblock {\em Phys. Rev. Lett.}, 120:141103, 2018.

\bibitem{george2017deep}
D.~George and E.~A. Huerta.
\newblock Deep neural networks to enable real-time multimessenger astrophysics.
\newblock {\em Phys. Rev. D}, 97(4):044039, 2018.

\bibitem{Gebhard_2019}
T.~D. Gebhard, N.~Kilbertus, I.~Harry, and B.~Schölkopf.
\newblock Convolutional neural networks: A magic bullet for gravitational-wave
  detection?
\newblock {\em Phys. Rev. D}, 100(6), 2019.

\bibitem{George_2018}
D.~George and E.~A. Huerta.
\newblock {Deep Learning for real-time gravitational wave detection and
  parameter estimation: Results with Advanced LIGO data}.
\newblock {\em Phys. Lett. B}, 778:64–70, 2018.

\bibitem{Menendez-Vazquez:2020khz}
A.~Men\'endez-V\'azquez, M.~Kolstein, M.~Mart\'{\i}nez, and Ll.~M. Mir.
\newblock {Searches for compact binary coalescence events using neural networks
  in the LIGO/Virgo second observation period}.
\newblock {\em Phys. Rev. D}, 103(6):062004, 2021.

\bibitem{PhysRevD.107.082003}
M.~Andr\'es-Carcasona, A.~Men\'endez-V\'azquez, M.~Mart\'{\i}nez, and Ll.~M.
  Mir.
\newblock {Searches for mass-asymmetric compact binary coalescence events using
  neural networks in the LIGO/Virgo third observation period}.
\newblock {\em Phys. Rev. D}, 107:082003, 2023.

\bibitem{Razzano_2018}
M.~Razzano and E.~Cuoco.
\newblock Image-based deep learning for classification of noise transients in
  gravitational wave detectors.
\newblock {\em Class. Quantum Grav.}, 35(9):095016, 2018.

\bibitem{Biswas_2013}
R.~Biswas et~al.
\newblock Application of machine learning algorithms to the study of noise
  artifacts in gravitational-wave data.
\newblock {\em Phys. Rev. D}, 88(6), 2013.

\bibitem{Cavaglia_2019}
M.~Cavaglia, K.~Staats, and T.~Gill.
\newblock {Finding the Origin of Noise Transients in LIGO Data with Machine
  Learning}.
\newblock {\em Commun. Comput. Phys.}, 25(4), 2019.

\bibitem{10.1093/mnras/stad3448}
M.~Andr\'es-Carcasona, M.~Mart\'{\i}nez, and Ll.~M. Mir.
\newblock Fast bayesian gravitational wave parameter estimation using
  convolutional neural networks.
\newblock {\em Monthly Notices of the Royal Astronomical Society},
  527(2):2887--2894, 2023.

\bibitem{KAGRA:2023pio}
R.~Abbott et~al.
\newblock {Open Data from the Third Observing Run of LIGO, Virgo, KAGRA, and
  GEO}.
\newblock {\em Astrophys. J. Suppl.}, 267(2):29, 2023.

\bibitem{gwosc}
Gravitational wave open science center.
\newblock \url{https://gwosc.org/}.

\bibitem{Davis_2021}
D.~Davis et~al.
\newblock {LIGO detector characterization in the second and third observing
  runs}.
\newblock {\em Class. Quantum Grav.}, 38(13):135014, 2021.

\bibitem{Acernese_2023}
F.~Acernese et~al.
\newblock {Virgo detector characterization and data quality: results from the
  O3 run}.
\newblock {\em Class. Quantum Grav.}, 40(18):185006, 2023.

\bibitem{2021PhRvX..11b1053A}
R.~Abbott et~al.
\newblock {GWTC-2: Compact Binary Coalescences Observed by LIGO and Virgo
  during the First Half of the Third Observing Run}.
\newblock {\em Phys. Rev. X}, 11(2):021053, 2021.

\bibitem{PhysRevD.109.022001}
R.~Abbott et~al.
\newblock {GWTC-2.1: Deep extended catalog of compact binary coalescences
  observed by LIGO and Virgo during the first half of the third observing run}.
\newblock {\em Phys. Rev. D}, 109:022001, 2024.

\bibitem{PhysRevX.13.041039}
R.~Abbott et~al.
\newblock {GWTC-3: Compact Binary Coalescences Observed by LIGO and Virgo
  during the Second Part of the Third Observing Run}.
\newblock {\em Phys. Rev. X}, 13:041039, 2023.

\bibitem{Usman2016TheCoalescence}
S.~A. Usman et~al.
\newblock {The PyCBC search for gravitational waves from compact binary
  coalescence}.
\newblock {\em Class. Quantum Grav.}, 33(21):215004, 2016.

\bibitem{nitz2017detecting}
A.~H. Nitz, T.~Dent, T.~Dal~Canton, S.~Fairhurst, and D.~A. Brown.
\newblock {Detecting binary compact-object mergers with gravitational waves:
  Understanding and Improving the sensitivity of the PyCBC search}.
\newblock {\em The Astrophysical Journal}, 849(2):118, 2017.

\bibitem{nitz2020gwastro}
A.~H. Nitz et~al.
\newblock {gwastro/pycbc: PyCBC release v1. 16.11}.
\newblock {\em Zenodo}, 2020.

\bibitem{PhenomPv3}
S.~Khan, K.~Chatziioannou, M.~Hannam, and F.~Ohme.
\newblock Phenomenological model for the gravitational-wave signal from
  precessing binary black holes with two-spin effects.
\newblock {\em Phys. Rev. D}, 100(2), 2019.

\bibitem{LIGOScientific:2019hgc}
B.~P. Abbott et~al.
\newblock {A guide to LIGO\textendash{}Virgo detector noise and extraction of
  transient gravitational-wave signals}.
\newblock {\em Class. Quantum Grav.}, 37(5):055002, 2020.

\bibitem{QTransf}
J.~C. Brown.
\newblock {Calculation of a constant Q spectral transform}.
\newblock {\em The Journal of the Acoustical Society of America},
  89(1):425--434, 1991.

\bibitem{DBLP:journals/corr/HeZRS15}
K.~He, X.~Zhang, S.~Ren, and J.~Sun.
\newblock Deep residual learning for image recognition.
\newblock {\em CoRR}, abs/1512.03385, 2015.

\bibitem{Kingma2014Adam:Optimization}
D.~P. Kingma and J.~L. Ba.
\newblock {Adam: A Method for Stochastic Optimization}.
\newblock {\em 3rd International Conference on Learning Representations, ICLR
  2015 - Conference Track Proceedings}, 2014.

\bibitem{abadi2016tensorflow}
M.~Abadi et~al.
\newblock {TensorFlow: A System for Large-Scale Machine Learning}.
\newblock In {\em 12th USENIX symposium on operating systems design and
  implementation (OSDI 16)}, page 265, (2016).

\bibitem{Abbott2005SearchStars}
B.~P. Abbott et~al.
\newblock Search for gravitational waves from galactic and extra-galactic
  binary neutron stars.
\newblock {\em Phys. Rev. D}, 72(8):23, 2005.

\bibitem{Ade:2015xua}
P.~A.~R. Ade et~al.
\newblock {Planck 2015 results. XIII. Cosmological parameters}.
\newblock {\em Astron. Astrophys.}, 594:A13, 2016.

\bibitem{astropy:2018}
{Astropy Collaboration} and {Astropy Contributors}.
\newblock {The Astropy Project: Building an Open-science Project and Status of
  the v2.0 Core Package}.
\newblock {\em The Astronomical Journal}, 156(3):123, 2018.

\end{thebibliography}

%\begin{thebibliography}{99.}
%%\bibitem{cms} The ATLAS Collaboration, [arXiv:hep-ex/1108.4908], submitted to Phys. Lett. B.\\
%%The CMS Collaboration, [arXiv:hep-ex/1110.3226], submitted to JHEP. 
%\end{thebibliography}

% =====================
% APPENDIX
% =====================
\clearpage
\appendix
\section{CNN results}
\label{sec:appendix}
Table~\ref{tab:appendix} collects the obtained CNN discriminants and the corresponding FAR values for the O3 events. 
Only the 50 events with the three detectors in science mode have been considered. 
Results are provided for all the three CNNs corresponding to high-mass, intermediate-mass, and low-mass ranges, 
as discussed in the body of the text. 

\begin{table} [h]
\begin{center}
\begin{scriptsize}
\begin{tabular}{l c c c c c c}\hline
\multicolumn{7}{c}{Results CNN scan over O3 events}\\ \hline
Event&\multicolumn{3}{c}{Discriminant}&\multicolumn{3}{c}{FAR (yrs$^{-1}$)}\\ 
 &high-mass&low-mass&intermediate-mass&high-mass&low-mass&intermediate-mass\\
\hline%
GW190403\_051519&0.79&0.18&0.5&37.34&$>$100&$>$100\\%
GW190408\_181802&1.0&0.2&0.88&0.01&$>$100&$>$100\\%
GW190412\_053044&0.99&0.24&0.9&0.01&$>$100&$>$100\\%
GW190413\_052954&0.77&0.15&0.73&68.62&$>$100&$>$100\\%
GW190413\_134308&0.88&0.29&0.41&5.01&$>$100&$>$100\\%
GW190426\_190642&0.28&0.22&0.26&$>$100&$>$100&$>$100\\%
GW190503\_185404&0.99&0.3&0.76&0.01&$>$100&$>$100\\%
GW190512\_180714&0.79&0.18&0.8&39.74&$>$100&$>$100\\%
GW190513\_205428&1.0&0.14&0.76&0.01&$>$100&$>$100\\%
GW190517\_055101&0.96&0.16&0.43&0.19&$>$100&$>$100\\%
GW190519\_153544&1.0&0.18&0.4&0.01&$>$100&$>$100\\%
GW190521\_030229&1.0&0.15&0.49&0.01&$>$100&$>$100\\%
GW190602\_175927&0.78&0.17&0.67&53.14&$>$100&$>$100\\%
GW190701\_203306&1.0&0.15&0.49&0.01&$>$100&$>$100\\%
GW190706\_222641&1.0&0.22&0.47&0.01&$>$100&$>$100\\%
GW190720\_000836&0.07&0.14&0.58&$>$100&$>$100&$>$100\\%
GW190725\_174728&0.08&0.25&0.65&$>$100&$>$100&$>$100\\%
GW190727\_060333&1.0&0.16&0.76&0.01&$>$100&$>$100\\%
GW190728\_064510&0.17&0.54&0.97&$>$100&$>$100&5.035\\%
GW190803\_022701&0.85&0.16&0.35&10.01&$>$100&$>$100\\%
GW190805\_211137&0.81&0.2&0.46&26.52&$>$100&$>$100\\%
GW190828\_063405&1.0&0.24&0.64&0.01&$>$100&$>$100\\%
GW190828\_065509&0.17&0.2&0.46&$>$100&$>$100&$>$100\\%
GW190915\_235702&0.99&0.17&0.54&0.01&$>$100&$>$100\\%
GW190916\_200658&0.68&0.17&0.39&$>$100&$>$100&$>$100\\%
GW190917\_114630&0.05&0.21&0.2&$>$100&$>$100&$>$100\\%
GW190924\_021846&0.05&0.16&0.56&$>$100&$>$100&$>$100\\%
GW190926\_050336&0.75&0.2&0.2&$>$100&$>$100&$>$100\\%
GW190929\_012149&0.77&0.23&0.3&76.78&$>$100&$>$100\\%
GW191105\_143521&0.05&0.2&0.62&$>$100&$>$100&$>$100\\%
GW191113\_071753&0.07&0.19&0.33&$>$100&$>$100&$>$100\\%
GW191127\_050227&1.0&0.19&0.65&0.01&$>$100&$>$100\\%
GW191215\_223052&0.81&0.15&0.49&32.00&$>$100&$>$100\\%
GW191219\_163120&0.09&0.16&0.34&$>$100&$>$100&$>$100\\%
GW191230\_180458&0.98&0.17&0.63&0.51&$>$100&$>$100\\%
GW200115\_042309&0.08&0.17&0.23&$>$100&$>$100&$>$100\\%
GW200129\_065458&0.95&0.15&0.62&1.56&$>$100&$>$100\\%
GW200202\_154313&0.08&0.33&0.6&$>$100&$>$100&$>$100\\%
GW200208\_130117&0.99&0.27&0.38&0.22&$>$100&$>$100\\%
GW200208\_222617&0.51&0.22&0.44&$>$100&$>$100&$>$100\\%
GW200209\_085452&0.78&0.23&0.46&86.13&$>$100&$>$100\\%
GW200210\_092254&0.08&0.39&0.62&$>$100&$>$100&$>$100\\%
GW200216\_220804&0.47&0.17&0.47&$>$100&$>$100&$>$100\\%
GW200219\_094415&0.8&0.15&0.39&45.064&$>$100&$>$100\\%
GW200220\_061928&0.13&0.18&0.31&$>$100&$>$100&$>$100\\%
GW200224\_222234&1.0&0.25&0.95&0.01&$>$100&12.61\\%
GW200308\_173609&0.23&0.2&0.31&$>$100&$>$100&$>$100\\%
GW200311\_115853&1.0&0.18&0.94&0.01&$>$100&23.00\\%
GW200316\_215756&0.2&0.16&0.19&$>$100&$>$100&$>$100\\%
GW200322\_091133&0.17&0.14&0.48&$>$100&$>$100&$>$100\\%
\hline
\end{tabular}
\end{scriptsize}
\caption{\small
CNN discriminants and the corresponding FAR values for the O3 events with all the three detectors in science mode. 
}
\label{tab:appendix}
\end{center}
\end{table}

\end{document}